\documentclass[aps,pre,twocolumn,groupedaddress]{revtex4}
\usepackage{amssymb}
\usepackage{graphicx}
\usepackage{mathtools}
\usepackage{graphicx}
\usepackage{epstopdf}
\usepackage{appendix}
\usepackage[english]{babel}

\newcommand{\cP}{\mathcal{P}}

\newcommand{\laplace}{\mathcal{L}}
\newcommand{\hf}{\hat{f}}

\usepackage{todonotes}

\begin{document}


\title{Population Density Equations for Stochastic Processes with Memory Kernels}


\author{Yi Ming Lai}
\author{Marc de Kamps}
\affiliation{Institute for Artificial and Biological Computation \\ School of Computing \\ University of Leeds \\ LS2 9JT Leeds \\ United Kingdom}


\date{\today}

\begin{abstract}
We present a novel method for solving population density equations (PDEs) - a mean field technique describing homogeneous populations of uncoupled neurons - where the populations can be subject to non-Markov noise for arbitrary distributions of jump sizes. The method combines recent developments in two different disciplines that traditionally have had limited interaction: computational neuroscience and the theory of random networks. The method uses a geometric binning scheme, based on the method of characteristics, to capture the deterministic neurodynamics of the population, separating the
deterministic and stochastic process cleanly. We can independently vary the choice of the deterministic model and the model for the stochastic process, leading to a highly modular numerical solution strategy. We demonstrate this by replacing the Master equation implicit in many formulations of the PDE formalism, by a generalization called the generalized Montroll-Weiss equation - a recent result from random network theory - describing a random walker subject to transitions realized
by a non-Markovian process. We demonstrate the method for leaky- (LIF) and quadratic-integrate and fire (QIF) neurons subject to spike trains with Poisson and gamma distributed interspike intervals. We are able to model jump responses for both models accurately to both excitatory and inhibitory input under the assumption that all inputs are generated by one renewal process.

\end{abstract}


\maketitle

\section{Introduction}

Population density techniques are widely used in physics, biology, chemistry, finance and other areas of science, often in the form of stochastic differential equation equations, or more generally in the form of the differential Chapman-Kolmogorov (dCK) equation \citep{gardiner1997}. The basic idea is always the same: the state of individuals in the population is described by a combination of deterministic laws that are known, and a noise process which is statistically similar for all individuals, causing irregular random state changes. 

Population density techniques have a long standing history in computational neuroscience starting with \citep{stein1965, johannesma1966,knight1972,gerstner1992}. 
In particular, the last twenty years have seen an explosion of interest in this area \citep{knight1996}, as it now becomes clear that although brain-sized simulations
are technically possible \cite{kunkel2011}, the resulting models are unwieldy, in terms of the number of parameters involved and the amount of data generated. Increasingly, the
population level is seen as an appropriate mesoscopic description level for modeling complex neural systems. For example, recently  a 
cortical column has been simulated  with 
population-based approaches, e.g.  \cite{cain2016,schwalger2017}. The development of techniques that relate the mesoscopic population level to that of individual neurons is 
therefore vital to the brain sciences.

In the past, many applications have used stochastic differential equations or alternatively  Fokker-Planck approaches: initially often for leaky-integrate-and-fire (LIF) neurons e.g. \citep{ricciardi,brunel1998}, but later also for other models such as quadratic- (QIF) or exponential-integrate-and-fire neurons \citep{brunel2003,fourcaud2003}, or even more complex ones such as the conductance-based model in \cite{richardson2004}. 
Many studies have assumed weak synaptic effects, allowing the introduction of a diffusion approximation and thereby the use of Fokker-Planck or Langevin equations. However, it has been argued that shot noise rather than Gaussian white noise is required for realistic simulations, for example by \citet{richardson2010} in the context of neocortical populations. Furthermore, post-synaptic effects are not necessarily small. Implicit in the formulation by \citet{omurtag2000} is the possibility that synaptic jumps are subject to Poisson statistics and may be large. Using  a similar framework \citet{nykamp2000} used a smoothness approximation for the population density that allows large synaptic inputs to be incorporated in a numerical approach. \citet{dekamps2013} and \citet{iyer2013} have demonstrated that by using the method of characteristics, a numerical scheme can be found for arbitrarily large jumps without relying on a smoothness assumption. 

By constructing a geometric binning scheme from the characteristics, we are able to model the deterministic neurodynamics by a shift of probability mass through the bins, 
thereby avoiding the numerical difficulties introduced by the drift term of the dCK equation. In this non-equidistant binning scheme, the full dCK equation is now reduced to a Poisson Master equation. 

This means that the system is represented by a combination of probability shifts and a Master equation which describes transitions due to a point process. In this paper, we relax the assumption that the noise is Poisson in nature. We can model the stochastic process as  a continuous-time random walk (CTRW) on a network of states, and follow the approach of \citet{hoffmann2013} to derive a generalized Montroll-Weiss equation; this leads to an equation analogous to the Poisson master equation, 
but with a convolution with a memory kernel based on the inter-arrival distribution of the point process. 

The importance of non-Poisson statistics has been pointed out by \citet{cateau2006}, who demonstrated that some experimental data is better described by a gamma distribution, and in a theoretical study showed that the dynamics of a synfire chain is substantially affected by the statistics of spike trains. Using a renewal-based approach, \citet{ly2009} were able to study non-Poisson inter-spike intervals by constructing a two-dimensional population density where one of the variables is the membrane voltage, and the other the time since the last spike. In this paper, we present a different approach, allowing us to create a general scheme suitable for different one-dimensional neuronal models for arbitrary transition matrices, thereby treating excitation and inhibition on the same footing, under the assumption that all transitions are generated by the same renewal process. Instead of a full two-dimensional treatment, we start with a one-dimensional method and find that the non-Markovian characteristics can be accounted for by a convolution with the recent history of the probability density of the population.  Other studies on the effect of non-Poisson noise, not directly related to the approach here, consider various forms of colored noise injected into individual neuron models  and studied the output statistics, for example  \citep{muller-hansen2015, shiau2015}.

\begin{figure}[h]

\includegraphics[width=.9\linewidth]{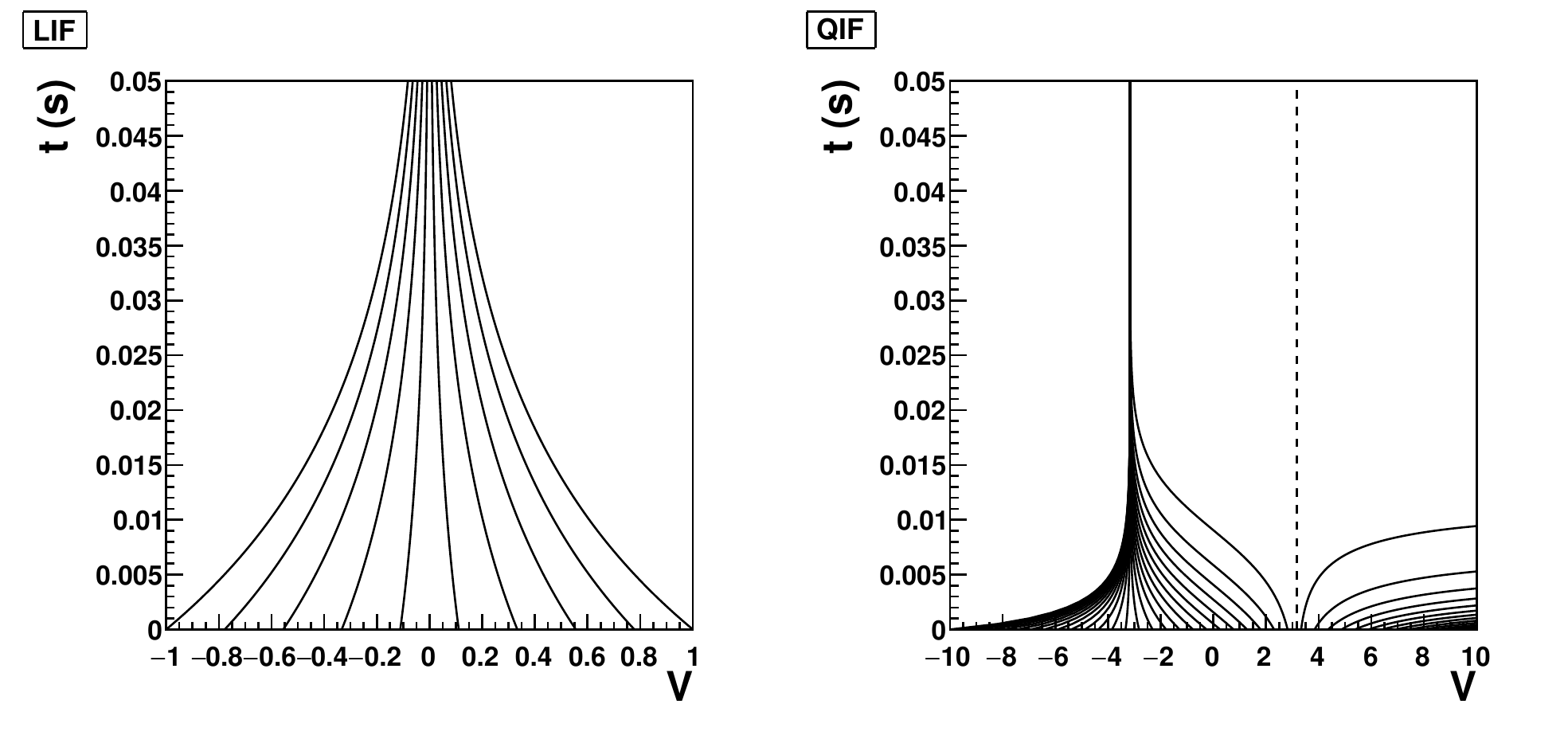}

\caption{Dynamics of LIF (left) and QIF neurons (right). Time $t$ is shown as a function of $V_0$ on an interval $[V_{min}, V_{max}]$.}
\label{fig-char}
\end{figure}

\section{Method}
We consider a population of neurons in the mean field approximation: in a homogeneous population neurons are uncoupled, identical, but individually see different realizations of the same statistical process.  Larger inhomogeneous networks must be described as homogeneous subpopulations and require assumptions on how the network connectivity transforms output of one population into the input of others, e.g. see \cite{amit1997}. Alternatively, we consider a single neuron subjected to a large number of repetitions of the same process. Under these assumptions  a population of individuals behaving according to $\dot{V} = F(V)$, with $V$ the membrane potential of a neuron, can be described by
the dCK equation (we will refer to the potential in lower case as an argument in the density and in upper case when discussing individual neurons for legibility):
\begin{equation}
  \frac{ \partial \rho }{ \partial t } + \frac{\partial}{ \partial v} ( F \rho ) = \int_{M} dw \left\{ W(v \mid w)\rho(w,t) - W(w \mid v  )\rho(v,t) \right\}  \, .
\label{eq-balance}
\end{equation} 

Here $\rho(v,t)$ is the population density defined on an interval $M$: $\rho(v)dv$ is the fraction of neurons with potential in $[v, v+ dv)$. $W(v \mid w)$ describes the transition density: the probability per unit time that a neuron moves from state $v$ to state $w$. $F(v)$ defines the neuron model. For example, the LIF neuron
is defined by $F(V) = -V/\tau$ where $\tau$ is the neuron time constant.  Other models include the QIF: 
\begin{equation}
F(V) = (V^2 + I)/\tau,
\end{equation} 
with $I$ often interpreted as a control parameter, or the exponential-integrate-and-fire model that we will not discuss here. The method here applies to any one-dimensional neural model. More complex neuronal models require more than one dimension; elsewhere we show that it is possible to apply the geometric binning scheme  to two dimensional neural models \cite{dekamps2016}. In this paper we will focus on one dimensional neural models as it allows a simpler exposition of the method.

\subsection{Geometric Binning}
Our objective is to describe the evolution of the density function $\rho(v,t)$. In the absence of synaptic input this is described by the advective part of the dCK equation, 
which could be solved numerically. However, geometrical considerations give a particularly simple method.
The non-dimensionalised LIF neuron is usually stated as:
\begin{equation}
  \tau \frac{dV}{dt} = -V,
  \label{eq-lifchar}
\end{equation}
where $V$ is the membrane potential, $\tau$ the membrane time constant of the neuron, so the membrane voltage decays exponentially in the absence of stochastic input. Explicitly, $t = \tau \ln \frac{V_0}{V}$ is the time it takes for the neuron to decay to a voltage $V$ from an initial voltage $V_0$. 

\begin{figure}
\includegraphics[width=0.8\linewidth]{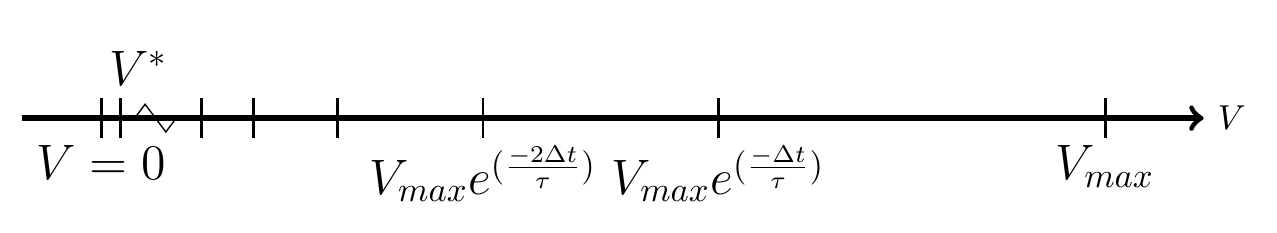}
\caption{A geometric grid for LIF neurons. (The bin $[0, V*]$ is a fiducial bin used to avoid having exponentially many bins near $V = 0$.) }
\label{fig-axis}
\end{figure}


We can use this to construct a set of characteristics (Fig. \ref{fig-char} left). If we consider a starting distribution of neurons in a population $\rho(V, t=0)$, it is clear that neurons between curves will remain  between those curves as time progresses,  in the absence of input. We can discretize  state space using equidistant steps in time, rather than potential: starting at $V_{max}$, we evolve Eq. \ref{eq-lifchar} during a time $\Delta t$, and use the new value of $V$ as a bin boundary. Repeating the process, we approach the equilibrium point $V =0$.  
This is shown diagrammatically in Fig. \ref{fig-axis}. The bins get exponentially smaller closer to $V = 0$, and therefore we define a small constant $V^*$ close to $V=0$.
We define a fiducial bin $[0, V^*]$ where probability mass remains stationary. Similarly, we approach the equilibrium from the left hand side by starting at $V_{min}$, and calculating the potential decay in steps of $\Delta t$, which yields a series of bin boundaries and break off in a similar way. In practice we are free to pick $V_{min}$, and usually pick a value that yields the same bin boundaries left and right of the equilibrium.

Starting from an arbitrary distribution of probability over the grid, its evolution can be done essentially without computation: each time step $\Delta t$, the mass in each bin - the fraction of the population present in that bin - moves to the next bin in the direction of the equilibrium. Mass that enters the equilibrium bin remains there. This simple observation suggests that problem can be solved by interleaving two steps: a shift of mass through a geometric grid, followed by a numerical solution of the Master equation which implements transport of mass from bin to bin. In the following section we prove this.

\begin{figure}[t]

\includegraphics[width=0.95\linewidth]{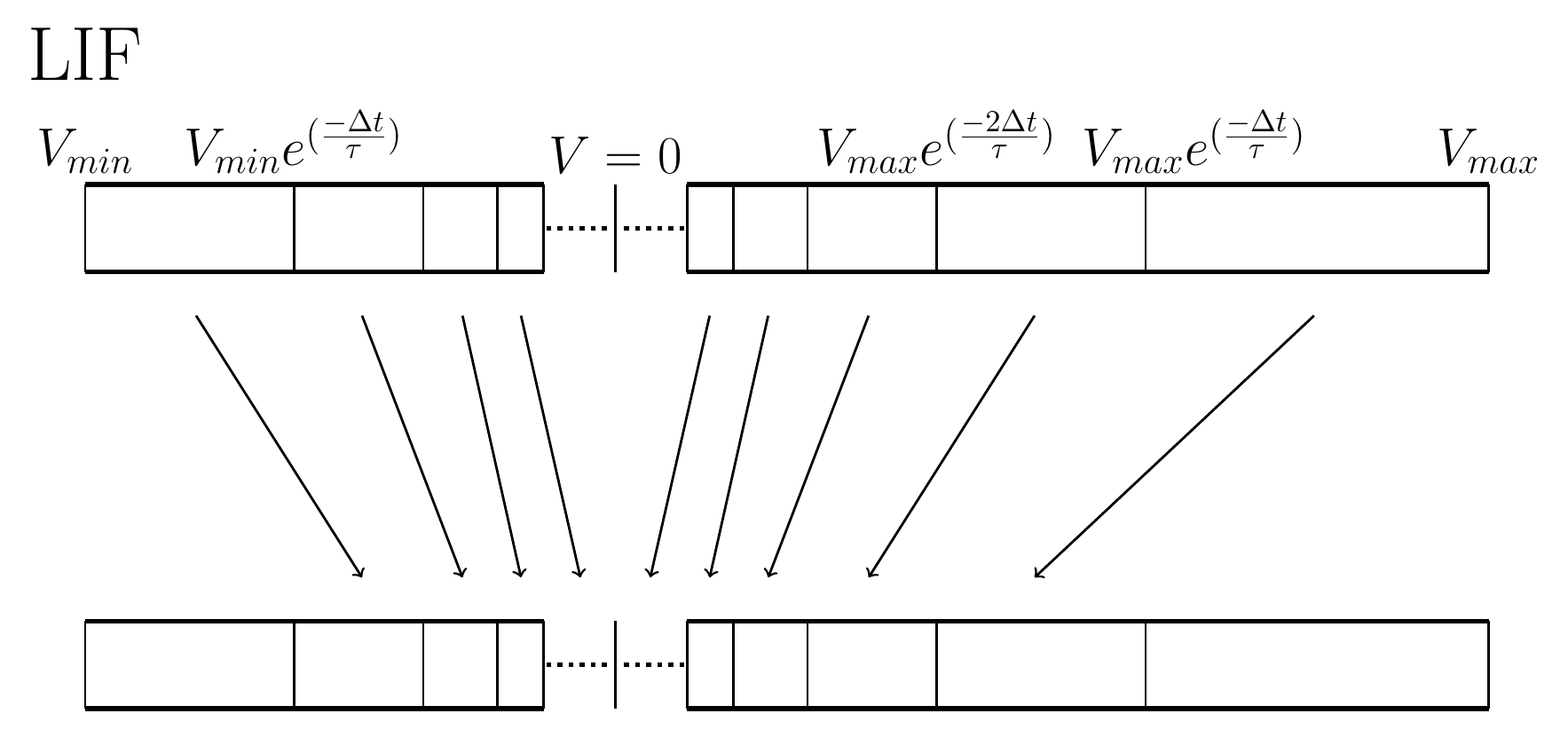} \\ \includegraphics[width=0.95\linewidth]{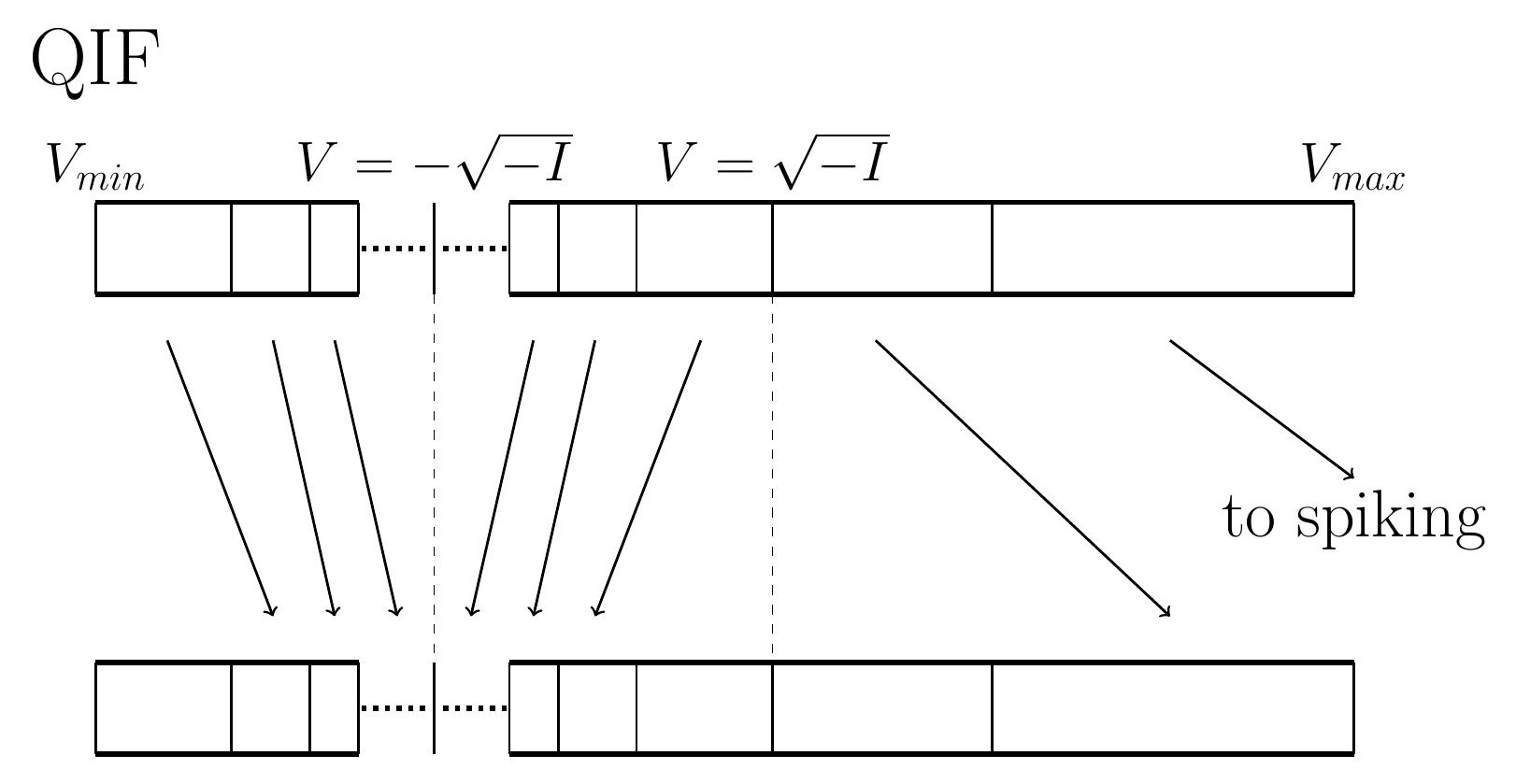} 

\caption{A shift of probability mass through the geometric grid is sufficient to capture the evolution of the density profile due to the deterministic neuronal
dynamics. For LIF neurons (top) mass moves in the direction of the equilibrium point. Mass that enters the fiducial bin surrounding the equilibrium point
remains there. For QIF neurons with $I<0$ (bottom), there are two equilibrium points: one stable, one unstable. Movement towards the stable equilibrium is similar to the
LIF case. Movement away from the unstable equilibrium towards the threshold is upwards. Probability mass reaching threshold must be removed from the system and
reinserted at a potential $V_{reset}$, possibly after observing a refractive period.}
\label{fig-schematic}
\end{figure}

\subsection{The Master Equation in a Moving Coordinate System}
 A key observation \cite{dekamps2003,ly2009,iyer2013,schwalger2015} is that the method of characteristics
can be used to transform Eq. \ref{eq-balance} into a simpler one.

Consider the ordinary differential equation:
\begin{equation}
\frac{d V}{dt} = F(V),
\label{eq-vv}
\end{equation}
and let $V(t, v(0))$ be a solution of Eq. (\ref{eq-vv}) with $v = v_0$ for $t = 0$ . It is possible to interpret this as a coordinate transformation:
\begin{equation}
  \begin{aligned}
    v^{\prime} &= V(t,v) \\
    t^{\prime} & =  t 
  \end{aligned}
\end{equation}

In the new coordinate system Eq. (\ref{eq-balance}) assumes a simpler form:
\begin{equation}
\frac{d \rho^{\prime}( v^{\prime}, t)}{dt} = \int_{M} dw \left\{ W(v^{\prime} \mid w)\rho(w) - W(w \mid v^{\prime}  )\rho(v^{\prime}) \right\},
\label{eq-simpler}
\end{equation}
with  
\begin{equation}
\rho^{\prime}(V(t),t) \equiv e^{\int^t_0 \frac{\partial F(V(\xi))}{\partial v} d \xi} \rho(V(t),t)
\label{eq-prime}
\end{equation}
This simpler form is explained by the observation that along integral
curves of the system, one can calculate the total derivative of the density
i.e. along curves $V(t)$ that are solution to Eq. (\ref{eq-vv}) we have 
$$
\frac{ d \rho(V(t), t)}{dt} = \frac{\partial \rho(V(t),t)}{\partial t} + \frac{ \partial \rho(V(t),t)}{\partial V}  \frac{ dV } {dt},
$$
using the chain rule. The definition of Eq. (\ref{eq-prime})
directly leads to Eq. (\ref{eq-simpler}).

Equation (\ref{eq-simpler}) is just the Master equation of the noise process, albeit in a moving coordinate system.
Intuitively, this makes sense: in a coordinate system that co-moves with the neuronal dynamics, all change must come from the stochastic process.
As an example, consider Poisson distributed spike trains. For shot noise:
\begin{equation}
W(w \mid v) = \nu \delta(w - v - h) + (1 - \nu) \delta(w -v),
\label{eq-shot}
\end{equation}
where $h$ is the synaptic efficacy and $\nu$ the rate of the Poisson process,
and $w, v$  arbitrary potential values. This indicates that the only possibility for a jump
is from a potential $v$ to $v+h$ as the transition probability is 0 for all other transitions.
The transition probability expresses that an  input spike causes an instantaneous jump in membrane potential.
For a Markov process $\nu$ and $h$ can be time dependent.

Consider the case of a LIF neuron, $F(V) = -V/\tau$. 
\begin{equation}
  \begin{aligned}
    v^{\prime} &= ve^{-\frac{t}{\tau}} \\
    t^{\prime} & =  t 
  \end{aligned}
\end{equation}

with $\rho^{\prime}(v^{\prime},t) = e^{-\frac{t}{\tau}}\rho(v^{\prime},t)$, and
Eq. (\ref{eq-balance}) reduces to:
\begin{equation}
\frac{\partial \rho}{\partial t} -\frac{1}{\tau}\frac{\partial}{\partial v}(\rho v) = \nu (\rho(v - h) -\rho(v)),
\label{eq-sharp}
\end{equation}
Of course a single synaptic efficacy is unrealistic and in practice one uses \cite{nykamp2000}:
\begin{equation}
\frac{\partial \rho}{\partial t} -\frac{1}{\tau}\frac{\partial}{\partial v}(\rho v) =\int dh p(h) \nu (\rho(v - h) -\rho(v)),
\label{eq-synapse}
\end{equation}
As we will argue below, this does not fundamentally change the method.

\begin{figure}[t]
\includegraphics[width=0.5\linewidth]{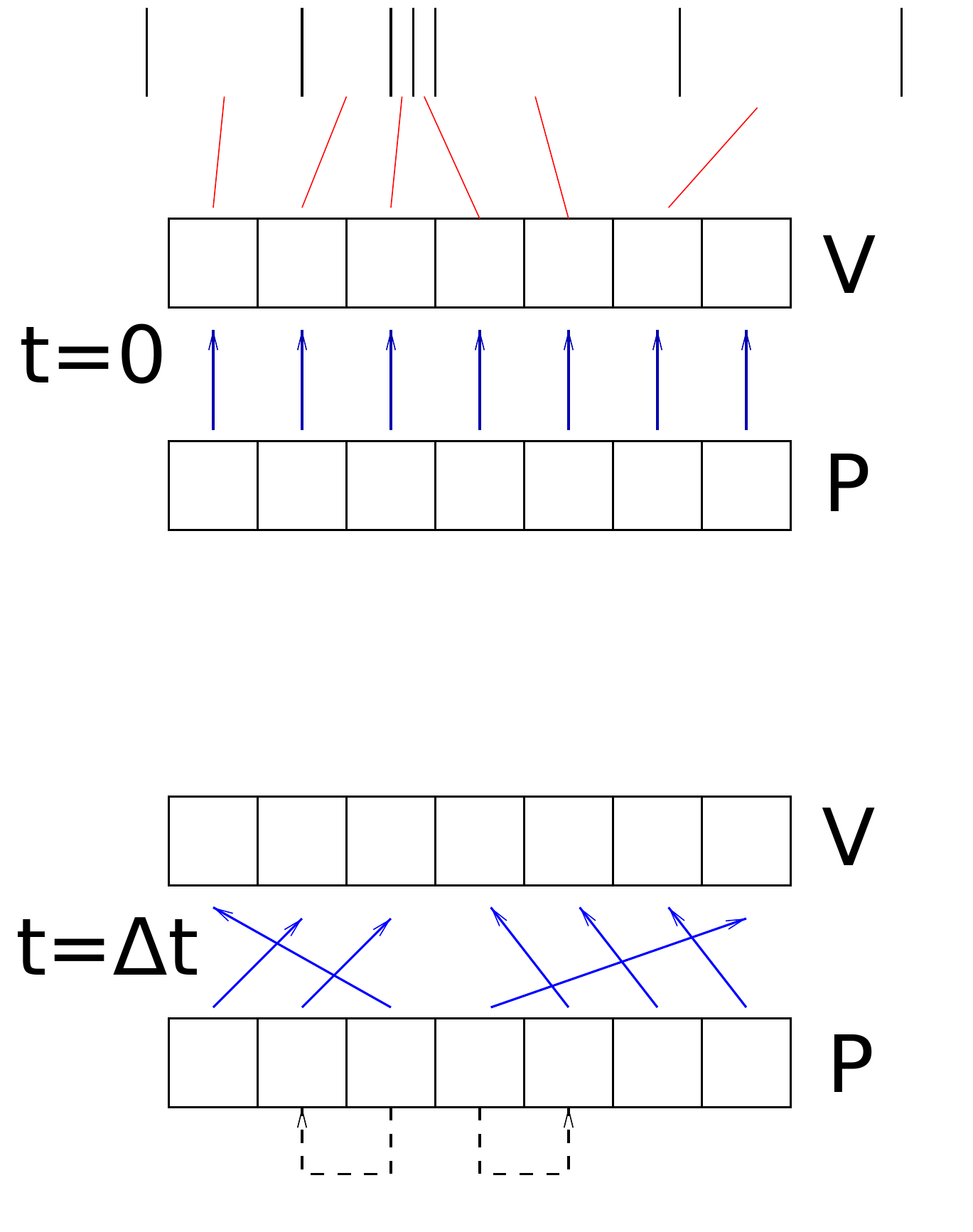}
\caption{The shift of probability mass can largely be replaced by an index update. The black bars at the top indicate a geometric grid. The array V stores the values of the
grid boundaries; the relationship between the contents of V and the grid boundaries are indicated symbolically by red lines. This relationship is immutable; it remains constant
throughout simulation. 
P is an array representing the probability mass at time $t$. The top figure represents the situation at $t=0$: each element of $P$ contains a numerical value
representing the amount of probability mass. For a given element, the blue arrow indicates the potential interval containing this; together the V and P arrays represent
a discretized density profile.  The evolution of the density profile is realized by updating the relationship between the P and the V array, as indicated in the bottom panel
representing the density profile at $t = \Delta t$, by the change of the blue arrows. The contents of the V array remain unchanged, as do the contents of P, with the exception
of two bins.  The situation depicted in this Figure shows decay towards a steady state that is represented by the third and fourth potential bin from the left.  Mass represented
by the two  outward pointing arrows on the extreme left and right represents mass that has cycled from the equilibrium bins to potential values at the extreme end of the potential
interval. If the stationary point is stable, this is undesirable and this mass should be removed and added to the elements of the mass array that currently point to the
equilibrium bins; this is indicated by the dashed arrows. This Figure shows a neural model that has a single stable equilibrium point, such as the LIF model. For a model
with more than one stationary point, such as the QIF neuron for  $I < 0$, the V and P arrays must be separated into isolated strips, each with their own relationship between
the P and V array. 
By updating the relationship between elements of P and V, the shift of mass is captured almost entirely  without  moving data around.}
\label{fig-shift}
\end{figure}

After the coordinate transformation this becomes:
\begin{equation}
\frac{d \rho^{\prime} (v^{\prime}, t)}{dt} = \nu(\rho^{\prime}(v^{\prime} - he^{\frac{t}{\tau}}) -\rho^{\prime}(v^{\prime})),
\label{eq-masterlif}
\end{equation}
where we have taken into account that Eq. (\ref{eq-shot}) must now be represented in $v^{\prime}$-space.
This constitutes a considerable simplification: instead of solving partial differential equation, one is faced with a system of ordinary differential equations. This
comes at a price: one is forced to  represent the density not in a fixed interval in potential space, but in a frame that moves with respect to that interval. Moreover,
as Eq. (\ref{eq-masterlif}) shows, in that  frame the jumps are time dependent, even if they are constant in the original frame. This precludes the analytic solution for constant $h$ given in \cite{dekamps2006}.

The geometric binning scheme provides a method for representing density in $v^{\prime}$ coordinates. The entire method now becomes a two step process. The first step consists of a shift of probability mass, as explained in Fig. \ref{fig-schematic}, which represents the movement of neurons under the influence of deterministic dynamics during a time step $\Delta t$. The second is the solution of the Master equation over a time step $\Delta t$, small enough for $h e^{\frac{\Delta t}{\tau}}$ and $\nu(t)$ to be constant. In the following section we will describe this process in detail.

Although shown for LIF neurons, the method generalizes in an obvious way. It is always the case that the characteristics of Eq. \ref{eq-balance} are given by the solutions
of the system $\tau \frac{dV} {dt} = F(V)$, and therefore a geometric grid can always be constructed by integrating this equation regardless of whether analytic solutions are 
available, like for QIF neurons, or a numerical solution is required. The only subtlety that needs to be observed is there may be multiple equilibria present; mass movement may be in opposite directions at either side of the equilibrium point. This is illustrated in Fig. \ref{fig-schematic}: LIF neurons have a single stable equilibrium point, QIF neurons a stable and unstable one (see Fig. \ref{fig-char}). It is best to think of a potential interval bounded by two equilibrium points (or the minimum or maximum potential) as an independent strip, and capture probability mass movement in each strip independently. The full potential interval is then represented by a collection of these
strips.

Finally, a point that is implementationally important. Rather than shifting the data around as described, which is computationally expensive, it makes more sense to
keep track of the position of each portion of probability mass in the geometric grid. This reflects the observation that the density profile is constant in $v^{\prime}$ space. 
The process is shown in Fig. \ref{fig-shift}.

Algorithmically, this introduces a considerable amount of bookkeeping, which is described in some detail in \cite{dekamps2013},
but which we will ignore in the remainder of the paper as it is not conceptually different from the method as described above.

\subsection{The Master Equation in a Geometric Grid}
\begin{figure}[h]
\includegraphics[width=1\linewidth]{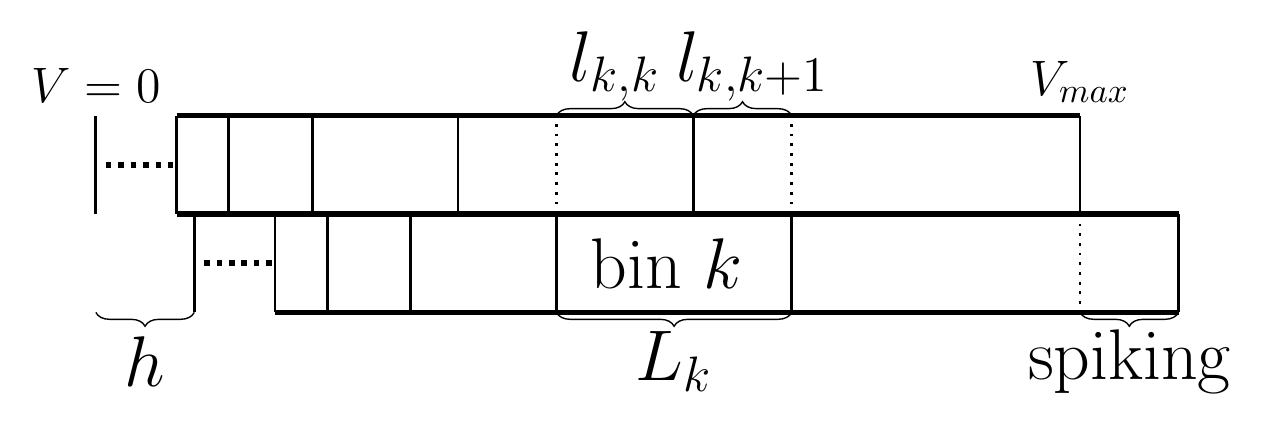} 
\caption{Coefficients for the Poisson Master equation are purely determined by synaptic efficacy $h$ and the bin boundaries. Two grids are shown above, with the bottom displaced by $h$. The transition matrix is determined by what fraction the displaced grid covers each bin of the original grid. For example, the bin labelled ``bin $k$'' has length $L_k$. After displacement by $h$, it overlaps bin $k$ by some interval $l_{k,k}$ and bin $k+1$ by an interval $l_{k,k+1}$. Therefore the transition matrix would have entries $m_{k,k} = \frac{l_{k,k}}{L_k} \,, m_{k,k+1} = \frac{l_{k,k+1}}{L_k}\,,$ and so on. At the same time, part of bin $k+1$ is pushed above threshold, which corresponds to spiking. The probability mass pushed above threshold is placed in the bin containing $V_{reset}$ at the next time step (or after a delay if a refractory period is desired). This procedure for bin $k$ allows us to calculate the $k$-th row of the transition matrix, so repeating the procedure for all bins gives us the full matrix. } 
\label{fig-trans}
\end{figure}

For simplicity, we will describe the solution to the Master equation of the Poisson process. Extension to the gMW equation will be straightforward. First, consider the Master equation at $t=k \Delta t$. We need to formulate the Master equation in a non-equidistant grid. Consider the probability mass in bin $i$. This bin corresponds to  a potential interval $[V_k(i), V_k(i+1)]$. Neurons that are present in this mass bin will, when they receive an input spike, move to a different potential and will be in the interval $[V_k(i) + h, V_k(i+1) + h]$. It is therefore a matter of finding out which potential intervals are covered by this interval, and by what proportion. This is a straightforward geometrical problem which is illustrated in Fig. \ref{fig-trans}.  

Denote the set of mass bins covered by $[V_k(i) + h, V_k(i+1) + h ]$ by $V_{k,i}(h)$ and for bin $j \in  V_{k,i}(h)$ let $m_{ij}$ denote what proportion of bin $j$ is covered by
$[V_k(i) + h, V_k(i+1) + h ]$ - then the Master equation becomes:
\begin{equation}
  \frac{ d \mathcal{P}[i] }{dt} = \nu \{ -\mathcal{P}[i] + \sum_{j \in V_{k,i}(h)} m_{ij} \mathcal{P}[j] \},
\label{eq-m}
\end{equation}
or in vector-matrix notation:
\begin{equation}
\frac{d\mathcal{P}}{dt} = \nu(-I + M)\mathcal{P},  
\label{eq-mastermat}
\end{equation}
where the elements of $M$ are $m_{ij}$. The vector $\mathcal{P}$ is the probability mass in our non-equidistant bins and corresponds to a discrete version of the quantity $\rho'(V) dV$ from the previous section. The process is identical for excitatory ($h > 0$) or inhibitory ($h < 0$) input.

Once we have computed this transition matrix $M$, we know the probability of an event causing a transition from any state to another.  Therefore the full implementation of our method consists of two interleaved steps: a probability shift between bins of the probability array, corresponding to the deterministic neuronal dynamics, while the  effect of the stochastic input is captured by solving the Master equation (\ref{eq-mastermat}) on the corresponding non-equidistant binning scheme. 

The matrices $M$ will be band matrices, i.e. sparse (see Fig. \ref{fig-matrix} for an explicit example). A row typically reflects a position in the interval from where the neurons leave and a position where the neurons
arrive, with the intermediate positions filled with zeros. Synaptic smearing broadens the band of the arrivals, 
but as long as the width of the synaptic distribution is small compared
to the simulation interval, the overall matrix $M$ will still be sparse. In practice, one samples the synaptic distribution with a few well chosen synaptic efficacies, yielding a number of matrices - the overall matrix $M$ is then a weighted sum of these matrices. As this is done before simulation starts, there is only a small effect on simulation time.

\section{Generalized Montroll-Weiss Equation: beyond Markov}

The master equation, when standing on its own, describes the behaviour of a random walker on a network, where each interval in $v$-space is a node. The walker is locked on a node, unless a connection to another node appears at which point the walker must move instantaneously after which the connection vanishes. In the context of computational neuroscience, the appearance of a connection is the arrival of an input spike, which allows the receiving neuron to move from its current membrane potential to a different one. The probability of a connection appearing is given by the previously-calculated transition matrix. 

Having used this abstraction, we now are able to extend our master equation method to other renewal processes by using a generalized Montroll-Weiss (gMW) equation for this network. The Montroll-Weiss equation was originally used to model anomalous diffusion on regular lattices, and was recently generalized to networks by \citet{hoffmann2013}. We now briefly restate the derivation of generalized Montroll-Weiss equation for our example, following the approaches of \citep{hoffmann2013, montroll1965, kenkre1973}. We start with a random walker in our state space with a waiting time distribution (WTD) $f(t)$. 

We are interested in determining $\cP(t) = \{P_0(t),P_1(t),\ldots,P_N(t) \}$ the probability that the walker will be at any given state at a time $t$. We define $q_i(t)$ to be the probability that a walker arrives at a state $i$ at exactly $t$, and $q_i^k(t)$ to be the probability that a walker arrives at state $i$ at time $t$ having taken exactly $k$ steps. (Hence $q_i(t) = \sum_{k=0}^\infty q_i^k(t)$.) If we know the $q_i^k(t)$s, we know that the probability of a walker being at state $j$ after $k+1$ steps is the sum of the $q_i^k(t)$s, weighted by the probability of making a step from $j$ to $i$ at the required time. Hence we construct the recursion relation:
\begin{equation}
q_j^{(k+1)}(t) = \int_0^t \sum_{\forall i} m_{ij} f(t-\tau) q_i^k(\tau) d\tau \,,
\end{equation}
where the $m_{ij}$s are the coefficients of the transition matrix induced by the synaptic efficacy $h$, as calculated in Fig. \ref{fig-trans}. In Laplace space:
\begin{equation}
\hat{q}_j^{(k+1)}(s) = \sum_{\forall i} m_{ij} \hf(s) \hat{q}_i^{k}(s) \,.
\end{equation}
Summing over all $k$ and adding $\hat{q}_j^0(s)$ to both sides gives:
\begin{equation}
\hat{q}_j^0(s) + \sum_{k=0}^\infty\hat{q}_j^{(k+1)}(s) = \sum_{k=0}^\infty  \sum_{\forall i} m_{ij} \hf(s) \hat{q}_i^k(s) + \hat{q}_j^0(s) \,,
\end{equation}
which in matrix-vector notation is 
\begin{equation}
\hat{q}(s) = M \hf(s)\hat{q}(s) + \hat{q}^0(s)\,.
\end{equation}
so that
\begin{equation}
\hat{q}(s) = \left(I - M\hf(s)\right)^{-1} \cP(0)\,.
\end{equation}
Now we know that the probability of a being at state $i$ at a time $t$ must be equal to the probability of arriving at state $i$ at some $\tau<t$ and an event not occurring between $\tau$ and $t$:
\begin{equation}
P_i(t) = \int_0^t G_i(t-\tau) q(\tau)d\tau\,,
\end{equation}
where $G_i(t) = 1 - \int_0^t f(t) dt$ is the probability of an event not occurring in time $t$. In Laplace space: $\hat{P}_i(s) = \hat{f}(s)\hat{q}_i(s)$, allowing us to use our expression for $q$ to give us the gMW equation for our network:
\begin{equation}
\hat{\cP}(s) = \frac{1-\hf(s)}{s} \left( I - M\hf(s) \right)^{-1} P(0)\,. 
\label{eq:laplacegmw}
\end{equation}
Next, we use the identity  $\laplace\{d\cP/dt\} = s\hat{\cP}(s) - P(0)$ and substitute in $P(0) = \frac{s}{1-\hf(s)}(1- M\hf(s))\cP(s)$ from Eq. (\ref{eq:laplacegmw}). Some rearrangement yields:
\begin{eqnarray}
\laplace\{\frac{d\cP}{dt}\} &=& (M-I) \hat{K} \hat{\cP}(s)\,\\
\frac{d\cP}{dt} &=& (M-I) \left(K(t) \ast \cP(t)\right) \,,
\end{eqnarray}
where the function $K$ is the memory kernel and is defined so that 
\begin{equation}
\hat{K}(s) \coloneqq \frac{s\hf(s)}{1-\hf(s)} \,. \label{eq:kernel}
\end{equation}
We note that this looks similar to our previous Master equation - in the case where we have a Poisson process of rate $\nu$, $K = \nu \delta(t)$ and we obtain our previous Master equation. 

In general $K$ does not have a closed-form solution and has to be evaluated numerically. For a gamma distribution of shape $\alpha$ and rate $\nu$, $K$ is the Laplace inverse of
\begin{equation*}
\hat{K} = \frac{s\nu^\alpha}{(s+\nu)^\alpha} - \nu^\alpha \,. 
\end{equation*}
In the interests of simplicity, in the subsequent examples we will consider $\alpha = 2$ and $3$, giving
\begin{eqnarray}
K(t)\mid_{\alpha=2} &=& \nu^2 \exp(-2\nu t) \nonumber \\
K(t)\mid_{\alpha=3} &=& \frac{2\sqrt{3}}{3}\nu^2\exp(-\frac{3}{2}\nu t)\sin(\frac{\sqrt{3}}{2}\nu t) 
\label{eq-kernel}
\end{eqnarray}
respectively. This is motivated by experimental data showing gamma-distributed inter-spike intervals \citep{cateau2006,VNS:4608056}. However, we stress that our method also works on other distributions for which the memory kernel has to be numerically evaluated. Efficient computation of the Laplace transform of other ubiquitous probability distributions of inter-event statistics of renewal processes, such as the Weibull or Pareto distributions, is an open area of research, for example \citep{rossberg2008}. 

We will compare the results of simulations between Poisson input and gamma input. In order to do this, we define $K_{norm}:=K/\int_0^\infty K(t) dt$, so that $K_{norm} \ast P$ is also a probability distribution.   This simplifies the equation in some cases, for example, if the population reaches a steady-state distribution $P_s$, then $K_{norm}\ast P_s = P_s$. In our examples, the normalisation constant $\int_0^\infty K(t) dt$ is equal to the expectation value 
of the input spike train (which is $\nu/\alpha$ for a $\Gamma(\alpha,\nu)$ distribution). Hence we can cast the integro-differential gMW equation for gamma input in the form:
\begin{equation}
\frac{d\cP}{dt} = \frac{\nu}{\alpha} (M-I) \left(K_{norm}(t) \ast \cP(t)\right) \,, 
\label{eq-gmw}
\end{equation}
allowing us to compare gamma distributions with different shapes by varying the rate $\nu$ accordingly. For other distributions, this comparison cannot always be done; for example, the method is also suitable for evaluating inputs with a power-law distribution which does not necessarily have finite moments.

\section{Results}

We consider our population of LIF neurons to have a membrane time constant of $\tau=0.05$ s, and begin with a single Poisson input of 800Hz with synaptic efficacy of 0.03, 
which has been used as a benchmark in earlier studies \citep{omurtag2000}. We then take the natural extension to gamma distributed inputs, and verify our method against Monte 
Carlo simulations. In the figures here, the initial condition is that all the neurons in the population are at their equilibrium $V=0$, we have normalized the threshold potential
$V_{th}$ so that $V_{th} =1$ and dimensionless. 

\begin{figure}[h!]
%
\includegraphics[width=.5\textwidth]{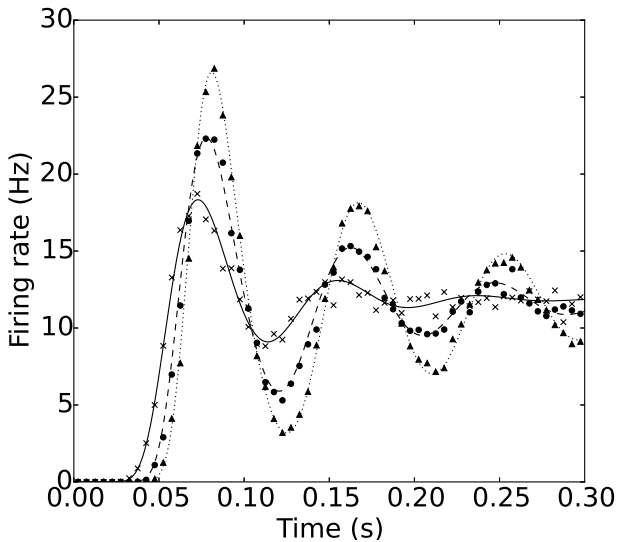}
%
\caption{Firing rates of the LIF neuron with inputs from a $\Gamma(\alpha,\nu)$ distribution. Lines are calculated using our method, while markers are from Monte Carlo simulations of 10000 neurons. $h=0.03, V_{th} = 1$ in all cases. Solid line and crosses: $\alpha=1, \nu=800$, i.e. a Poisson process with rate 800. Dashed line and circles: $\alpha=2, \nu=1600$. Dotted line and triangles: $\alpha=3, \nu=2400$. ($\nu$ is varied such that the expectation of the input process remains the same across all cases.)} 
\label{fig-transient}
\end{figure}

In Fig. \ref{fig-transient} we  observe good agreement with Monte Carlo simulations in the firing rate. We note that our method works much faster, as the computational load scales approximately linearly with the number of bins in our discretized characteristic space (which does not depend on the system size), while the Monte Carlo simulations scale with the number of neurons. In this paper we use on the order of tens of thousands of neurons in our Monte Carlo simulations. 

\begin{figure}[h!]

\includegraphics[width=.45\textwidth]{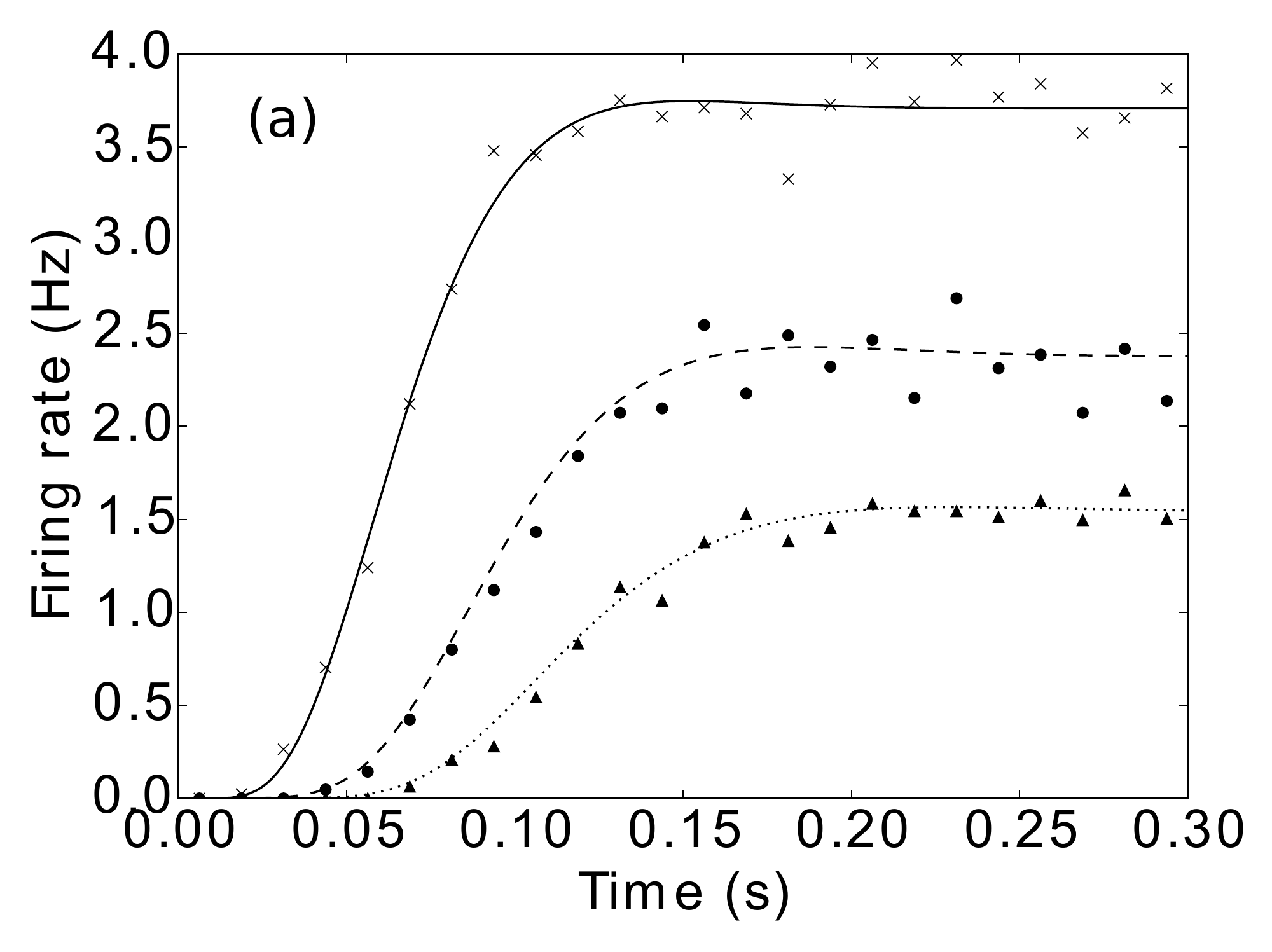}
\includegraphics[width=.45\textwidth]{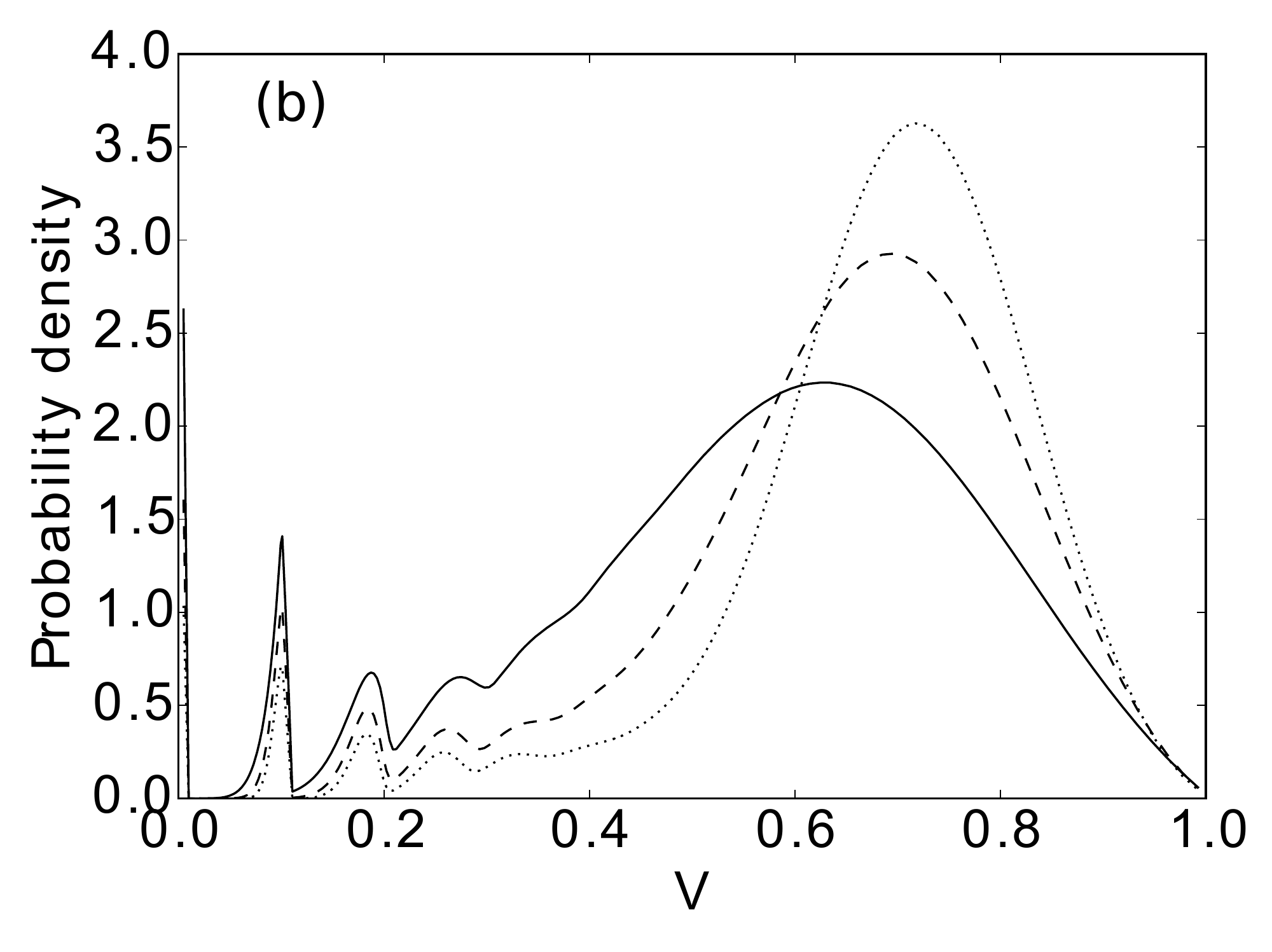}

\caption{(a): Firing rates of the LIF neuron with inputs from a $\Gamma(\alpha,\nu)$ distribution. Lines are calculated using our method, while markers are from Monte Carlo simulations of 10000 neurons. $h=0.1$ in all cases. Solid line and crosses: $\alpha=1, \nu=150$, i.e. a Poisson process with rate 150. Dashed line and circles: $\alpha=2, \nu=300$. Dotted line and triangles: $\alpha=3, \nu=450$. ($\nu$ is varied such that the expectation of the input process remains the same across all cases.) (b): the steady state density profiles for the different shape factors.}
\label{fig-lowsteady}
\end{figure}

We also see that for higher shape factors, the population experiences stronger transients and takes longer to reach its steady state firing rate. In  Fig. \ref{fig-lowsteady}, we show that changing the shape factor of the input distribution can even change the steady-state firing rate in low firing rate regimes. The density profiles
are also significantly affected by the shape factor. 

\begin{figure}[h!]
\includegraphics[width=.45\textwidth]{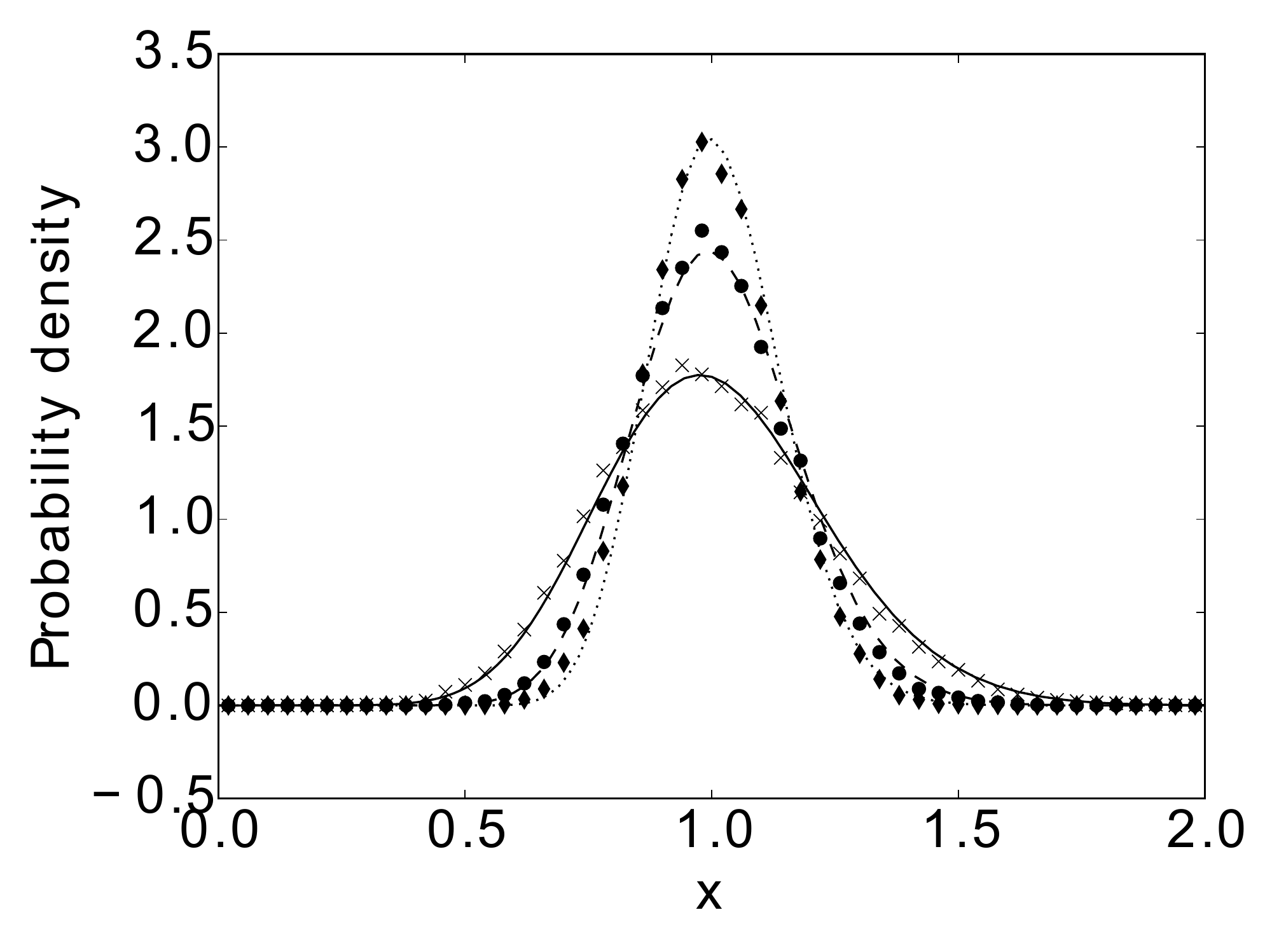}
\caption{Steady-state density of the generalised OU process. Lines are calculated using our method, while markers are from Monte Carlo simulations of 20000 neurons. $h=0.1$ in all cases. Solid line and crosses: $\alpha=1, \nu=10$, i.e. a Poisson process with rate 10. Dashed line and circles: $\alpha=2, \nu=20$. Dotted line and diamonds: $\alpha=3, \nu=30$.}
\label{fig-OU_fig}
\end{figure}

We contrast this with a system without threshold, where we see that decreased shape factor results in a broader steady-state density distribution around the same mean - see Fig. \ref{fig-OU_fig}. The system we consider is a generalization of the Ornstein-Uhlenbeck (OU) process. The OU process is one of the most fundamental examples of a stochastic process and is often used as a canonical example when developing techniques in the study of stochastic differential equations (SDEs) in various fields (\citep{doob1942,JOFI:JOFI4011,0026-1394-45-6-S17}). It is often written as: $dx_t = \theta (\mu-x_t) dt + \sigma dW_t$, where $W_t$ is the Wiener process. Here we replace the Wiener process jump process with an arbitrary probability density function for the time between jumps. In the absence of noise, the variable $x_t$ relaxes to $\mu$ with a time constant $\theta$. We consider a dimensionless version where $dx/dt=-x$ between jumps. For the stochastic part, we consider the variable $x$ to have jumps of size $h$ with the interval between jumps distributed according to the gamma distribution with shape $\alpha$ and rate $\nu$. 

\begin{figure}[h!]
\includegraphics[width=0.45\textwidth]{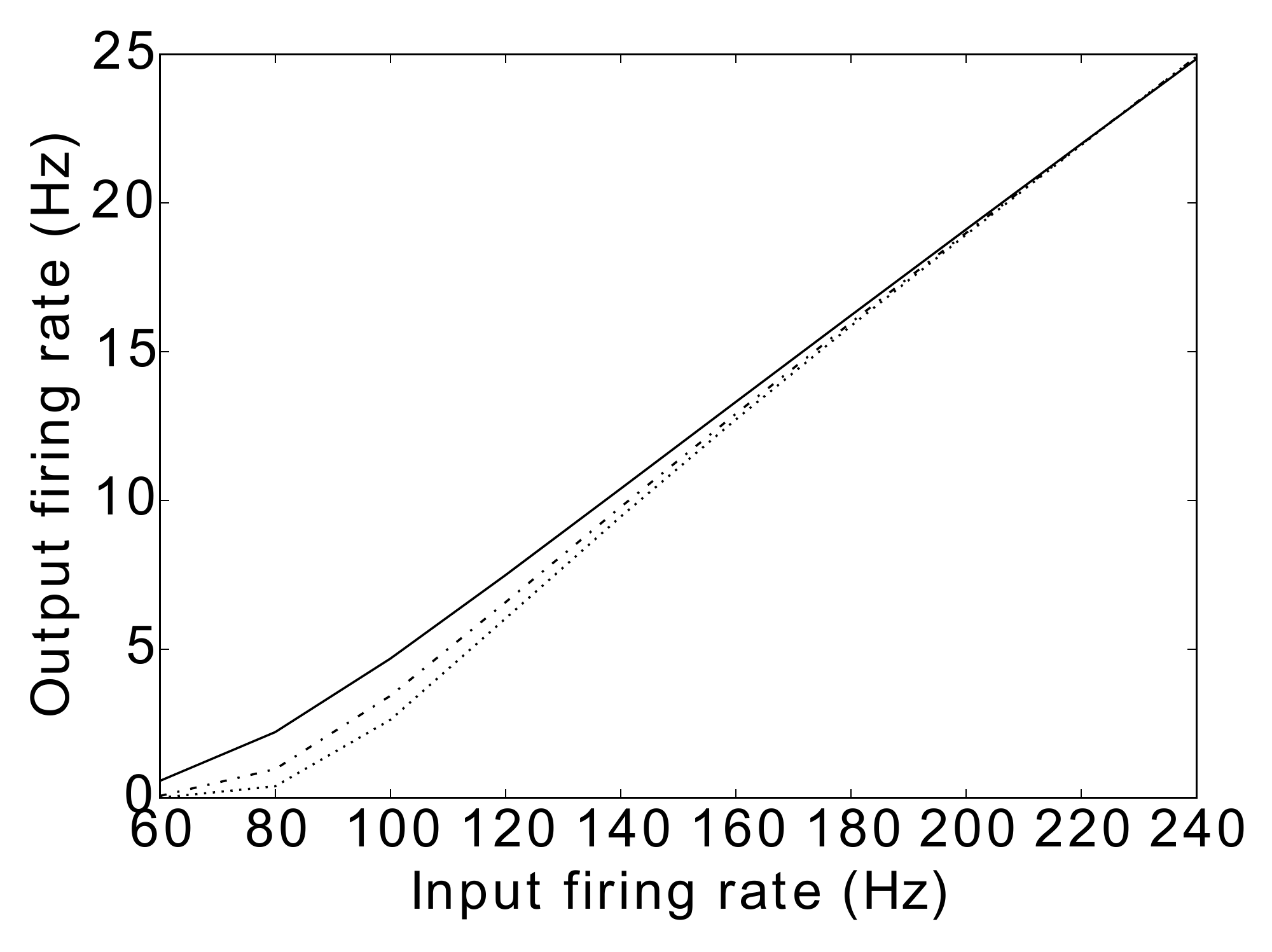}
\caption{Gain curves for $h=0.15$. Solid line: shape = 1 (Poisson process). Dashed line: shape = 2. Dotted line: shape = 3. Input firing rate is expected input $\nu/\alpha$. }
\label{fig-gain}
\end{figure}

Returning to our study of the LIF neuron, by considering a `gain curve' (Fig.  \ref{fig-gain}) of steady-state output firing rate against input firing rate for different inter-spike distributions, we can identify regions of parameter space where one would expect to see significant differences induced by different shape factors.
As we can see from the gain curves, for the same expected input and efficacy an increased shape factor decreases the firing rate. While this effect is only slight at high input firing rates, it is significant at lower firing rates, and we can see it changes the threshold input required for firing. 

\begin{figure}[h!]
\includegraphics[width=0.45\textwidth]{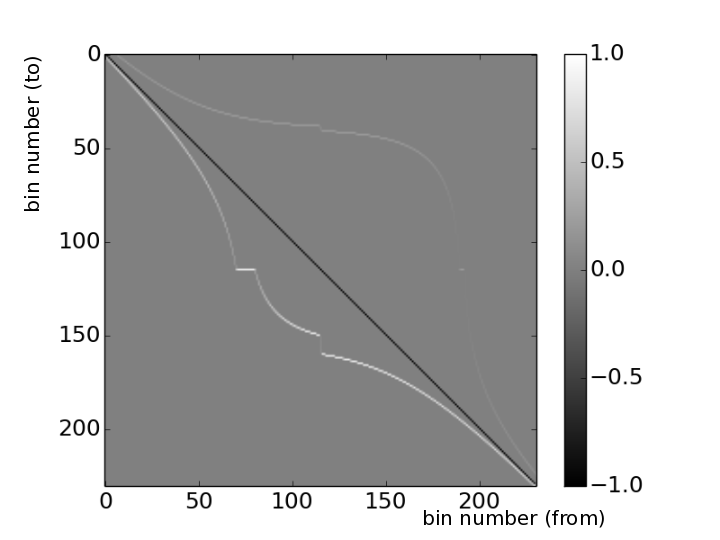}
\caption{The transition matrix $\mathcal{M}$ as an operator for moving probability mass from bin $i$ to bin $j$. 
Synaptic noise removes neurons from their current position, resulting in a loss term along the diagonal. Neurons undergoing an excitatory jump move up in potential
and thereby end up in the mass array with a higher bin number. Neurons undergoing inhibition end up at a lower bin number, at a larger distance from the diagonal,
reflecting $h_i = -4 h_e$.  The complex shape of the two bands is a result of using a geometric grid: near
the reversal potential the same jump in potential covers more bins. The reversal bin is larger than the neighboring geometric bins, so upon translation
covers a large number of them. This represents the straight part of the bands. }
\label{fig-matrix}
\end{figure}

\begin{figure}[h!]
\includegraphics[width=.45\textwidth]{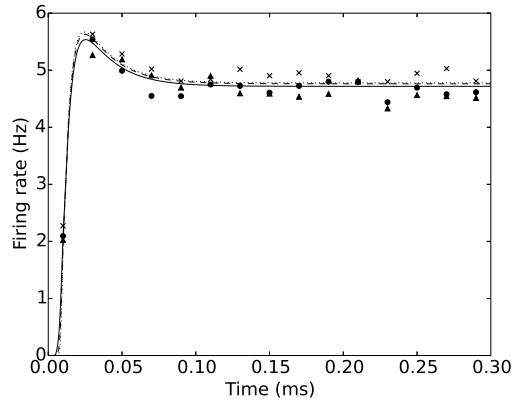}
\caption{Firing rates of the LIF neuron with inputs from a $\Gamma(\alpha,\nu)$ distribution. Lines are calculated using our method, while markers are from Monte Carlo simulations of 10000 neurons. In all cases, an input spike has an 0.8 probability of being excitatory ($h=0.05$) and an 0.2 probability of being inhibitory ($h=-0.2$). Solid line and crosses: $\alpha=1, \nu=2000$, i.e. a Poisson process with rate 2000. Dashed line and circles: $\alpha=2, \nu=4000$. Dotted line and triangles: $\alpha=3, \nu=6000$. ($\nu$ is varied such that the expectation, $\nu/\alpha$, of the input process remains the same across all cases.)}
\label{fig-balanced}
\end{figure}

By using the assumption by \citet{cateau2006} that a neuron experiences a superposition of many spike trains with little 
connectivity between them, so that the conglomerate spike train can be modeled as a single renewal process, we can study the balance of excitation and inhibition 
(Fig. \ref{fig-balanced}).  In the vein of previous studies such as \cite{amit1997}, we consider a 4:1 ratio of excitatory to inhibitory input, 
and a corresponding 1:4 ratio of synaptic efficacy. 
We generate a spike train with gamma distributed interspike intervals for different shape factors with a given
input rate $\nu$ as a marked point process. We perform a Bernoulli trial on each spike to determine whether it is excitatory ($p_e = 0.8$), 
or inhibitory ($p_i = 1-p_e$). An excitatory spike will contribute an instantaneous jump of magnitude $h_e = 0.05$ to the membrane potential while an inhibitory 
spike contributes a jump of magnitude
$h_i = -4h_e$. The resulting Monte Carlo simulations are given in Fig. \ref{fig-balanced}. To construct a population density version of this process,
we generated two matrices $\mathcal{M}_e$ and $\mathcal{M}_i$ by the process outlined in Fig. \ref{fig-trans} and add them to obtain a single matrix $\mathcal{M} \equiv p_e\mathcal{M}_e + p_i \mathcal{M}_i$, whose structure we represent visually in Fig. \ref{fig-matrix}. Using discretized versions of the kernel Eq. (\ref{eq-kernel}) then
allows us to solve Eq. (\ref{eq-gmw}) numerically for this case. Again, we interleave solutions over a time $\Delta t$ with the mass shift procedure to obtain the results
of Fig. \ref{fig-balanced} (solid curves). There is good agreement with the Monte Carlo process, and both methods predict a small output firing rate,
which is variability driven, given that the expectation value of the input contribution is 0. Surprisingly, we see no discernible dependency on the shape factor here.

\begin{figure}[h!]
\includegraphics[width=0.45\textwidth]{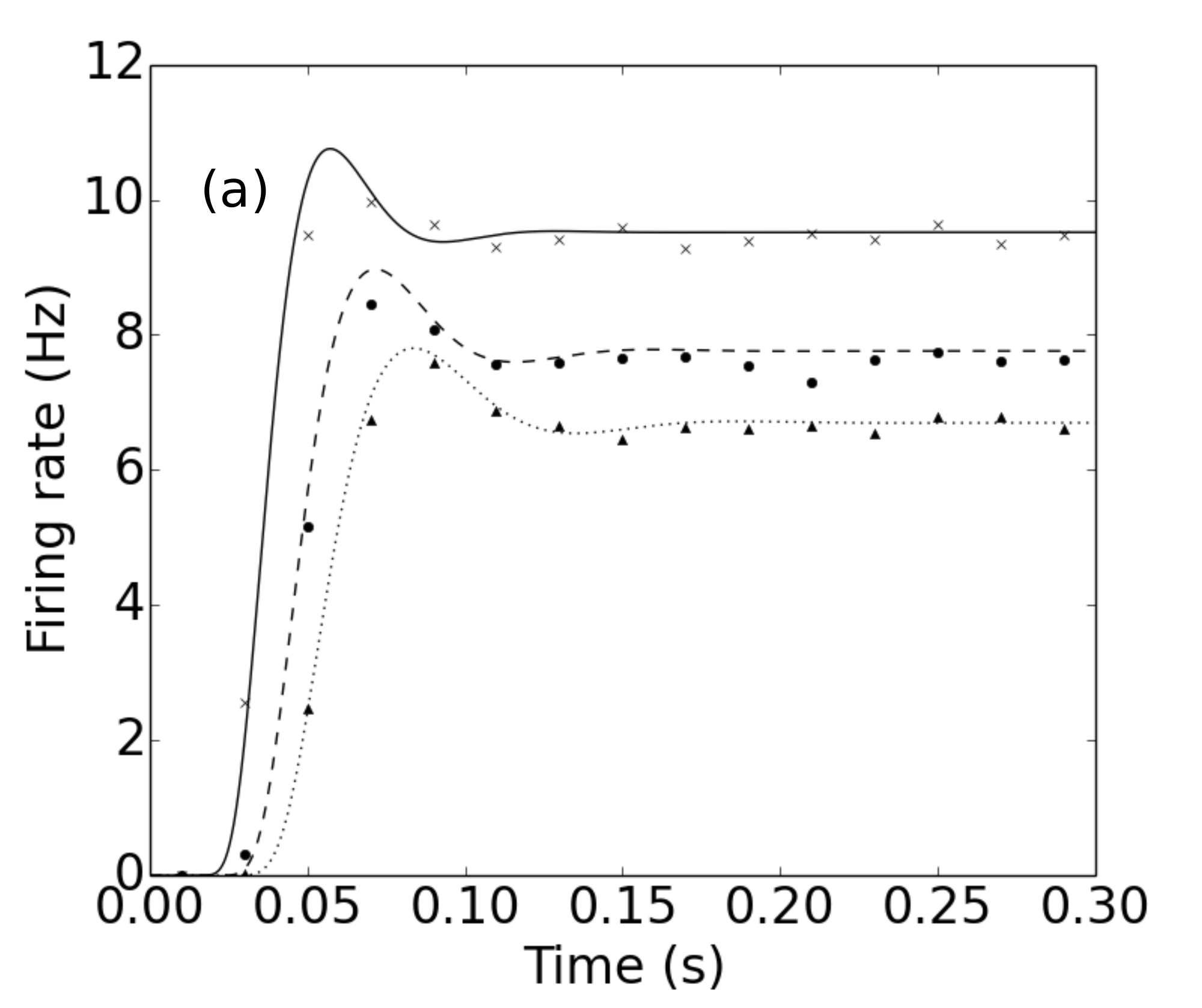}
\includegraphics[width=0.45\textwidth]{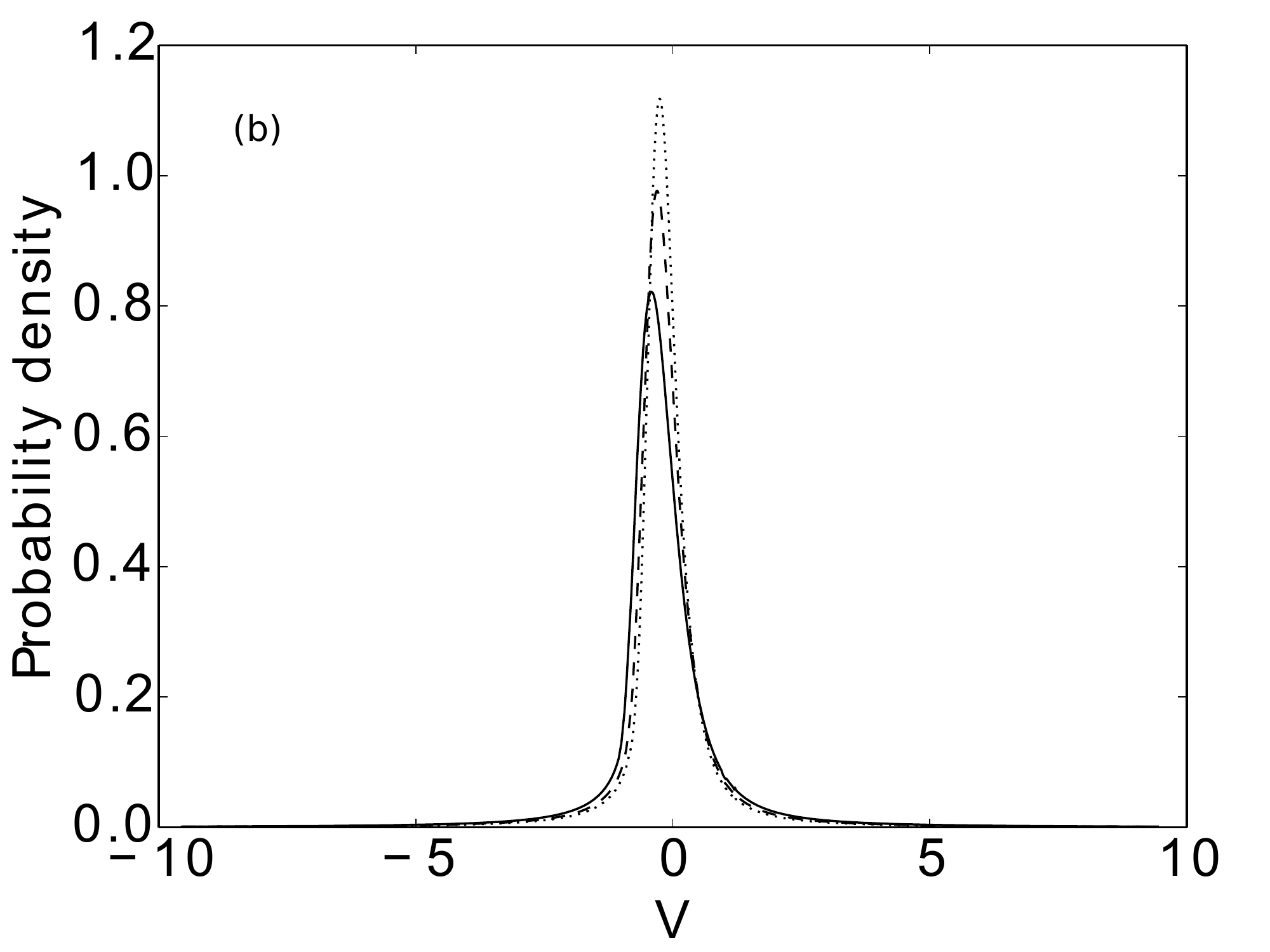}

\caption{(a): Firing rates of the QIF neuron with inputs from a $\Gamma(\alpha,\nu)$ distribution. Lines are calculated using our method, while markers are from Monte Carlo simulations of 10000 neurons. The population has a time constant $\tau = 0.01$ and a constant current $I = -1$, and the stochastic input has a synaptic efficacy $h=0.2$ in all cases. Solid line and crosses: $\alpha=1, \nu=500$, i.e. a Poisson process with rate 500. Dashed line and circles: $\alpha=2, \nu=1000$. Dotted line and triangles: $\alpha=3, \nu=1500$. ($\nu$ is varied such that the expectation of the input process remains the same across all cases.) (b): the steady state
density profiles for the different shape factors. The membrane potential has been renormalized so the $V_{th} = 1.0$ and is dimensionless.}
\label{fig-qif}
\end{figure}

Finally, we can easily obtain results from other neuronal models as well. In Fig. \ref{fig-qif} we show the steady state density profile of a population of QIF neurons ($I=-1)$, as well as the transient firing rate as response to a jump in input. Whilst the rate responses look qualitatively similar, the density profile looks different: neurons tend to cluster in the ghost of the attractor (the stable fixed point at $V = -1$). However, both the LIF (Fig. \ref{fig-lowsteady} (bottom)) and QIF cases display a shift in the peak of the probability density due to the shape factor. 

We attribute this shift in the peak to neurons returning to a lower potential value after having been pushed through threshold. Lower shape factors imply larger variability, and therefore more neurons being pushed across threshold. These neurons will reappear at the reset potential and move upwards in $V$, contributing to the density below the expectation value and a leftwards shift in the peak. We note that in the absence of a threshold no such shift is observed, as seen in Fig. \ref{fig-OU_fig}.

\subsection*{Multiple renewal processes}

A key assumption in our analysis so far is that our inputs, whether excitatory or inhibitory, can be assumed to be from a single conglomerate renewal process. 
Relaxing this assumption is difficult since superpositions of renewal processes are not themselves renewal processes (except the case where the component processes are 
Poisson \citep{samuels1974, ferreira2000}). \citet{hoffmann2013} derive a generalized Montroll-Weiss equation for a random walker on a network where transitions between nodes can 
be from different renewal processes. They do this by assuming that after a move is made by the random walker, the clocks of all renewal processes are reset. As such, only the 
joint probability distribution of the first event has to be used in the derivation, as opposed to a full description of a superposition of processes. 

However, in the context of a neuronal population receiving inputs from external sources or other populations, we cannot usually rely on this assumption. We briefly 
examine what occurs if we naively use the approach from \citep{hoffmann2013} to model a population receiving excitatory and inhibitory inputs, each of which is a process with 
inter-arrival times given by $\Gamma(2, 2\nu)$. In the case of a single conglomerate gamma process, this would give us a normalization constant $\int_0^{\infty} K(t) = \nu$, 
and for two inputs, one would expect a combined value of $2\nu$. However, when we compute the memory kernel for two inputs, we instead obtain 
$\int_0^{\infty} K(t) = {8\nu}/{5}$, i.e. a suppression to 80\% of the value that one would expect based on the individual processes. 

\begin{figure}
\includegraphics[width = .45\textwidth]{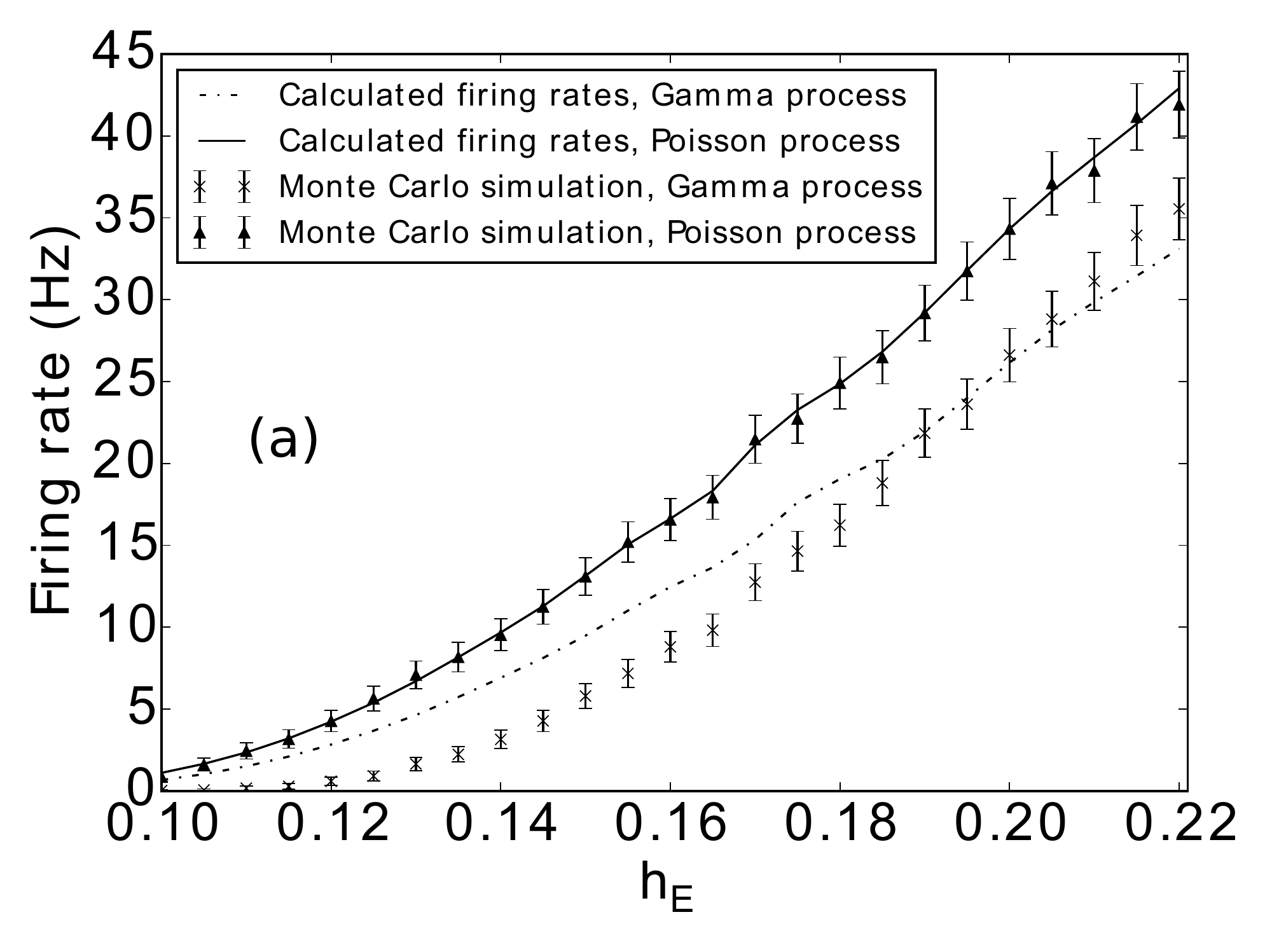}
\includegraphics[width = .45\textwidth]{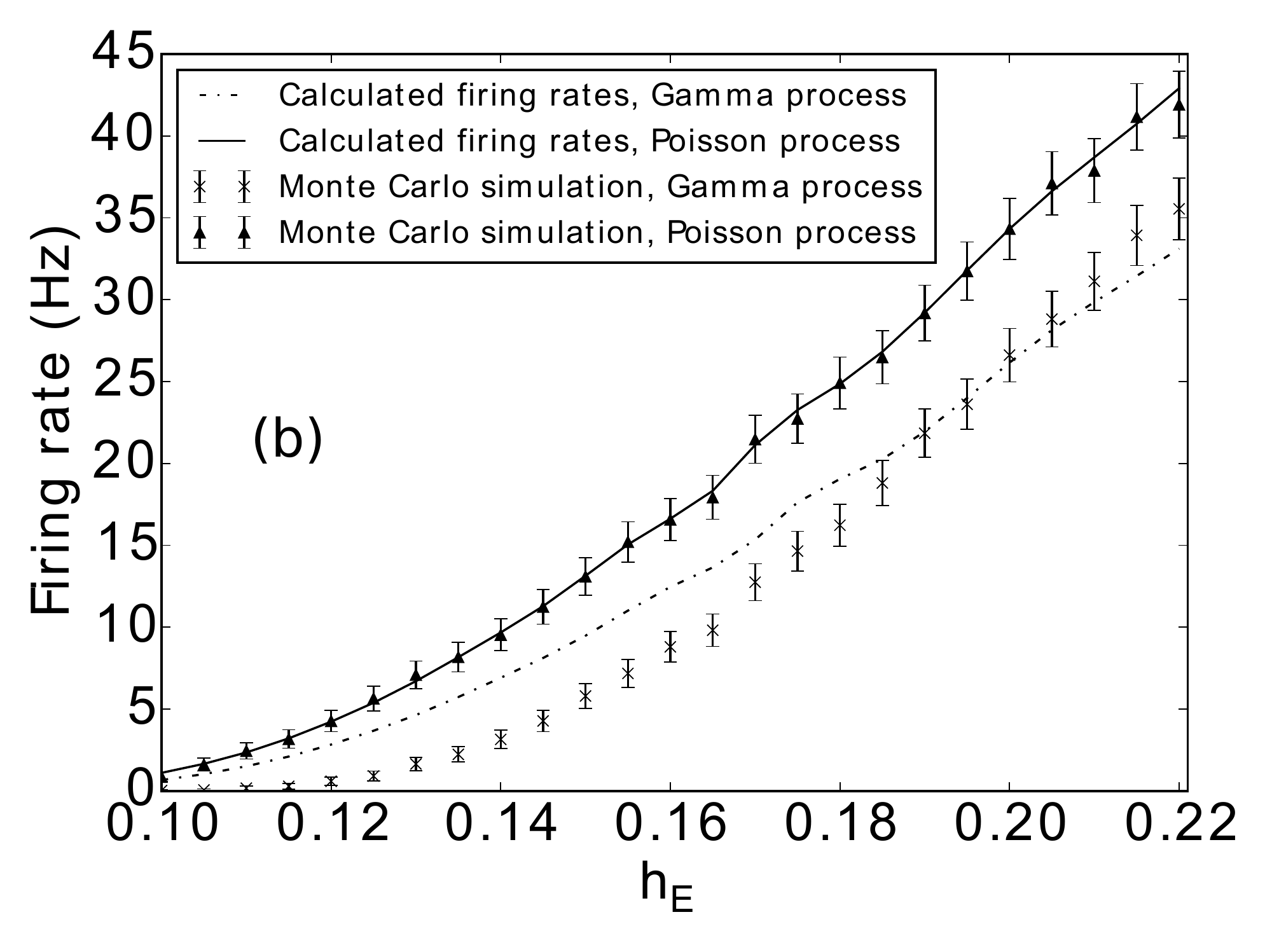}
\caption{The firing rate of a LIF population with two input processes, as a function of excitatory synaptic efficacy. The solid lines are the simulation of Poisson noise with our method, while the dash-dotted line is the solution of the Montroll-Weiss equation for the gamma processes.  The Monte Carlo simulations are the triangles and crosses with error bars. $\tau = 0.05$, $\nu = 500$. The inhibitory synaptic efficacy $h_I$ is fixed at 0.15.}
\label{fig-2ch}
\end{figure}
In Fig. \ref{fig-2ch} we examine the accuracy of this assumption. We see that in the Fig. \ref{fig-2ch} (a), when both processes make a comparable contribution, that there is reasonable 
agreement between the method and Monte Carlo simulations. However, where $h_e \gg h_i$, one would expect a convergence to the single channel result (i.e. where $\int_0^{\infty} K(t) = 2\nu$). In the Fig. \ref{fig-2ch} (b), we extend our regime to higher values of $h_e$, and indeed we see that Monte Carlo simulations approach the single channel result (which we label as ``theoretical correction''), while the gMW equation keeps predicting a reduction. This is because despite the relative insignificance of the inhibitory input spikes, they reset the clock for the excitatory process, leading to erroneous predictions.

\section{Conclusion and future work}
We have demonstrated a method for numerically solving population density equations that separates the deterministic and stochastic processes. The dynamics of the deterministic 
process are reflected in the choice of grid for the probability mass. Deterministic motion can be accounted for by shifting the mass through the grid. This just leaves the problem
of solving the equation  determining the mass transfer due to the stochastic process. For a Poisson process this is an extremely simple system of ordinary differential equations, 
with a resulting method that is manifestly insensitive to the gradient of the density profile.

The separation between deterministic and stochastic is general and makes no assumptions about the nature of the stochastic process. Therefore other methods for describing 
stochastic processes than Master equations can be incorporated. We demonstrated this explicitly by adopting a recent result from random network theory: the generalized
Montroll-Weiss equations. This leads immediately to a formulation of population density equations for stochastic processes with a memory kernel. 

Arbitrary synaptic distributions can be specified by choosing the appropriate transition matrices, and inhibition does not have to be considered as a separate special case - 
we can study systems with balanced excitation and inhibition. In general, one can model a synaptic distribution by using a superposition of transition matrices. Therefore, 
modeling learning can be easily accommodated for as it amounts to a reweighting of these matrices in the computation of the final transition matrix. As this final matrix is 
sparse, that can be done efficiently.

To summarize, we first construct a geometric binning using the method of characteristics, such that the deterministic dynamics of a neuron model can be captured by a probability 
shift from one bin to the next at each time step $\Delta t$. Between these steps we solve Eq. (\ref{eq-gmw}) numerically using the forward Euler method. This requires sampling 
the history at each time step; nevertheless, the resulting algorithm is still more efficient than Monte Carlo simulation. Furthermore, in most cases the memory kernel is of 
finite width, allowing further computational savings on the convolution and less memory of the process stored.

We note that Eq. (\ref{eq-gmw}) is of a simpler form than that in \citet{hoffmann2013}. There the authors construct a general method for a random walker on a network with 
arbitrary WTDs between nodes, whereas we consider that our input process has a single common WTD. This is due to the difference in the underlying assumptions - in their case they 
assume that the clocks of all WTDs are reset when the walker makes a move. On the other hand, from a neuroscience perspective, a neuron receiving an input spike should not affect 
the clocks of the neurons emitting said spikes.  A similar assumption was used in \citep{ly2009}. In this framework, 
we are able to consider distributions of synaptic efficacies, as well as mixed excitation and inhibition, as long as the conglomeration of spike trains can still be modeled as a 
renewal process. 

Dealing with the superposition of renewal processes in general is still an open problem in mathematics, as the superposition is in general no longer a renewal process:  it is a renewal process if and only if both processes are Poisson \citep{samuels1974, ferreira2000}. 

The method is not necessarily restricted to one dimensional neuronal models - as would be required for e.g. neurons displaying  a limit cycle. We have successfully implemented a method for two dimensional neural models \cite{dekamps2016}, and anticipate that there may be some additional computational overhead due to the need to retain a history of densities, but expect an otherwise straightforward generalization. 

We have briefly mentioned power-law distributions. One probability distribution that behaves as a power law asymptotically (for $0<\nu\leq 1)$ is the Mittag-Leffler 
distribution \citep{gorenflo2014}, which has $f^{ML}_\beta(t) = t^{\beta-1}E_{\beta,\beta}(-t^\beta)$, where 
$E_{\alpha,\beta}(z):= \sum_{n=0}^\infty z^n/\Gamma(\beta + \alpha n)$ is the generalized Mittag-Leffler function. This has the nice property that 
$\hf_\beta^{ML}(s) = (1+s^\beta)^{-1}$, so the memory kernel in our case would 
be $K = \laplace^{-1}[s^{1-\beta}] = t^{\beta - 2}/ \Gamma(\beta - 1)$. 
Hence the convolution $K \ast P = \int_0^t \frac{(t-\tau)^{\beta - 2}}{\Gamma(\beta - 1)} \cP(\tau) d\tau$, which is simply the Caputo fractional derivative 
\citep{caputo1967} $D^{1-\beta}_t \cP(t)$. Our gMW equation therefore becomes (by taking a fractional integral on both sides): 
\begin{equation}
D_t^\beta \cP(t) = (M - I) \cP(t) \,.
\end{equation} 
It would be interesting to explore the implications of this fractional differential equation for the population density. For example, there may be a connection with the model 
for adaptation posed by \citet{teka2014}, where a fractional derivative is introduced in the LIF model itself. 

As our method can be applied to any dynamical system with jump noise, we hope that our method is useful beyond computational neuroscience. An obvious application area is 
queuing theory, where the class of G/D/k queues handles events that arrive stochastically, but where the queues themselves operate deterministically.

The main limitations of our method are in studying superpositions of processes which cannot be approximated by conglomerate renewal process; and renewal processes with 
time-varying parameters, which is an important outstanding problem.

\section{Acknowledgment}
This project received funding from the European Union's Horizon 2020 research and innovation programme under grant agreement No. 720270 (Human Brain Project).

\bibliography{bib}

\end{document}